\documentclass[epj]{svjour}
\usepackage{graphicx}
\input{epsf}
\usepackage{amssymb}
\begin{document}

\title{Ulam method and fractal Weyl law for Perron--Frobenius operators}

\author{L.Ermann \and D.L.Shepelyansky}
\institute{Laboratoire de Physique Th\'eorique (IRSAMC), 
Universit\'e de Toulouse, UPS, F-31062 Toulouse, France
\and
LPT (IRSAMC), CNRS, F-31062 Toulouse, France
\and
http://www.quantware.ups-tlse.fr}

\date{February 4, 2010 }


\abstract{
We use the Ulam method to study spectral properties
of the Perron-Frobenius operators of dynamical maps
in a chaotic regime. For maps with absorption
we show that the  spectrum is characterized
by the fractal Weyl law recently established for
nonunitary operators describing poles of
quantum chaotic scattering with the Weyl exponent $\nu=d-1$,
where $d$ is the fractal dimension of corresponding strange set of trajectories nonescaping in future times.
In contrast, for dissipative maps we
find the Weyl exponent $\nu=d/2$ where $d$ is the 
fractal dimension of strange attractor. 
The Weyl exponent can be also expressed via the relation
$\nu=d_0/2$ where $d_0$ is the fractal dimension of the invariant sets.
We also discuss the
properties of eigenvalues and eigenvectors of such operators
characterized by the fractal Weyl law.}

\maketitle

\section{ Introduction}
The Weyl law gives a fundamental relation between a number of quantum states 
in a given classical phase space volume and an effective Planck constant 
$\hbar$ for Hermitian operators \cite{weyl}. Recently, this relation 
has been extended to nonunitary quantum operators which describe
complex spectrum of open systems or poles of scattering
problem. In this case the fractal Weyl law determines a number of 
Gamow eigenstates
in a complex plane of eigenvalues
with finite decay rates $\gamma$ via a fractal dimension
$d$ of a classical fractal set of nonescaping orbits.
The Gamow eigenstates find applications in various types 
of physical problems including 
decay of radioactive nuclei 
\cite{gamow}, quantum chemistry reactions \cite{moiseyev},
chaotic scattering \cite{gaspard} and chaotic microlasers \cite{stone1998}.
It is interesting that the fractal Weyl law was first introduced by
mathematicians via rigorous mathematical bounds  
\cite{sjostrand}. Later, numerical simulations for
systems with quantum chaotic scattering and open quantum maps
confirmed the mathematical bounds and determined a number of interesting
properties of such nonunitary quantum operators 
\cite{zworski2003,schomerus,keating,dls2008,saraceno2009}.
Open quantum maps with absorption, e.g.
the Chirikov standard map \cite{chirikov1979},
are very convenient for numerical studies that 
allowed to establish a number of intriguing 
properties of decay rates and quantum fractal eigenstates
in the limit of large matrix size and small scale quantum resolution
\cite{borgonovi,dls2008}.

\begin{figure}[ht]
\begin{center}  
\includegraphics[width=.225\textwidth]{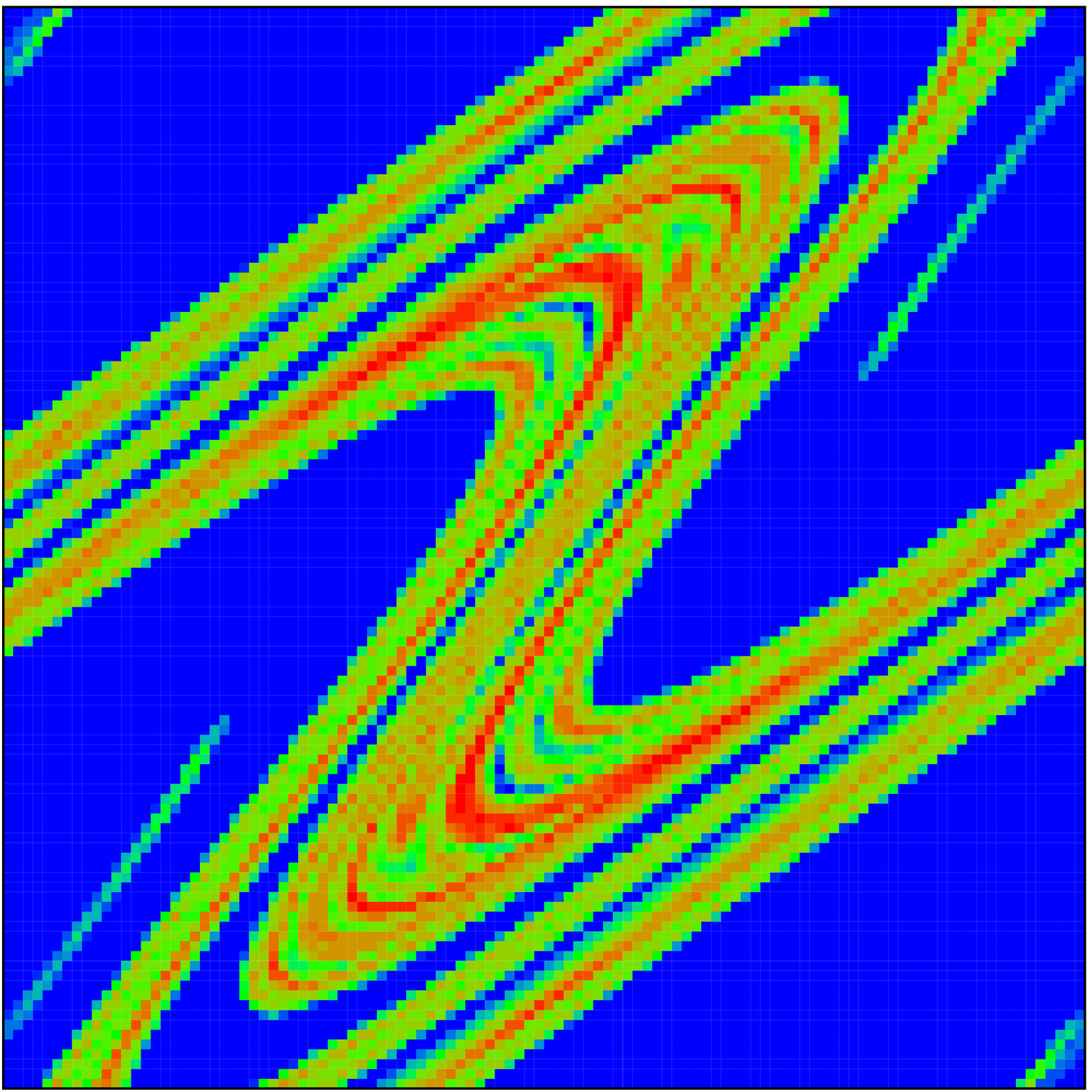} \hspace{0.01\textwidth}
\includegraphics[width=.225\textwidth]{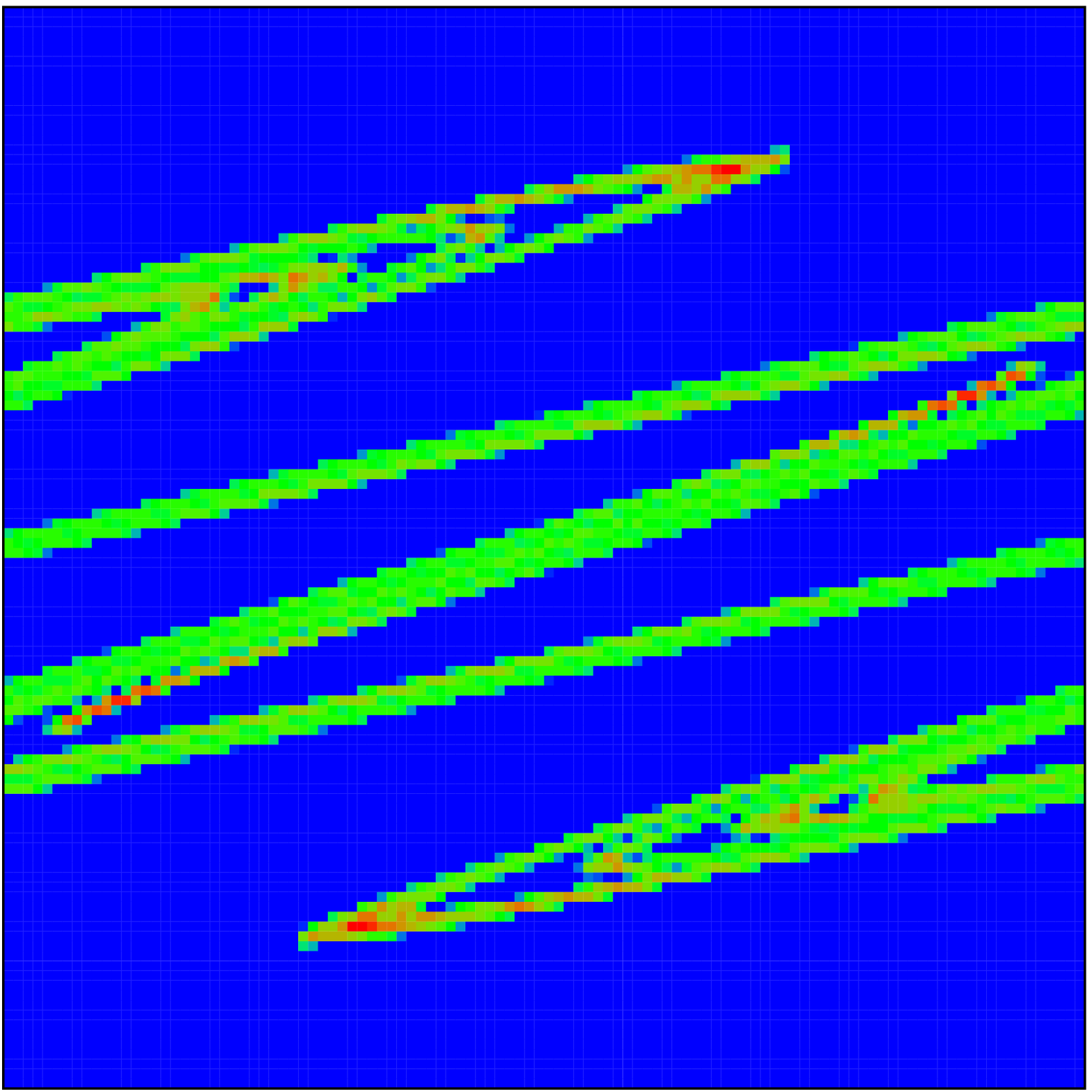} \\
\vspace{0.01\textwidth}
\includegraphics[width=.225\textwidth]{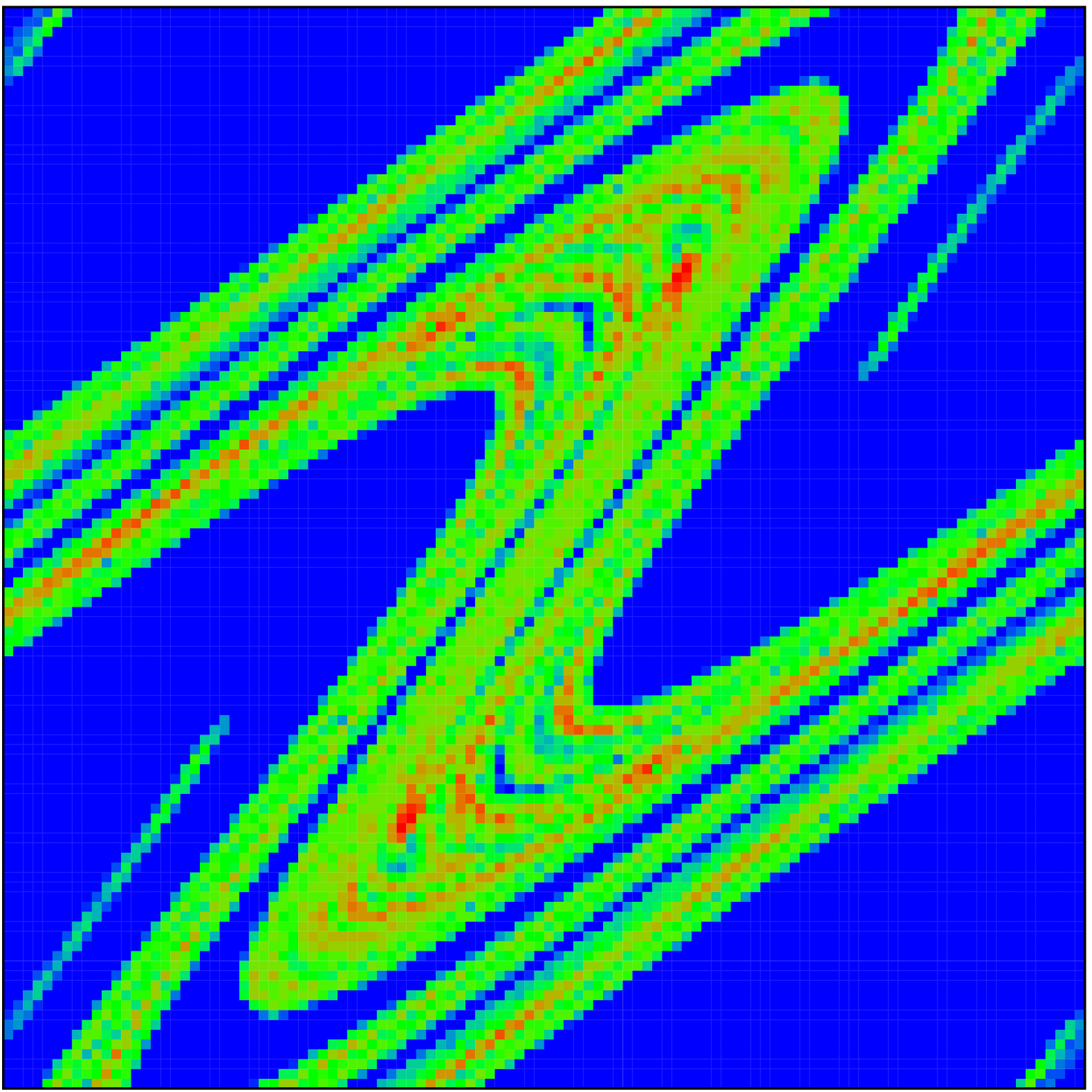} \hspace{0.01\textwidth}
\includegraphics[width=.225\textwidth]{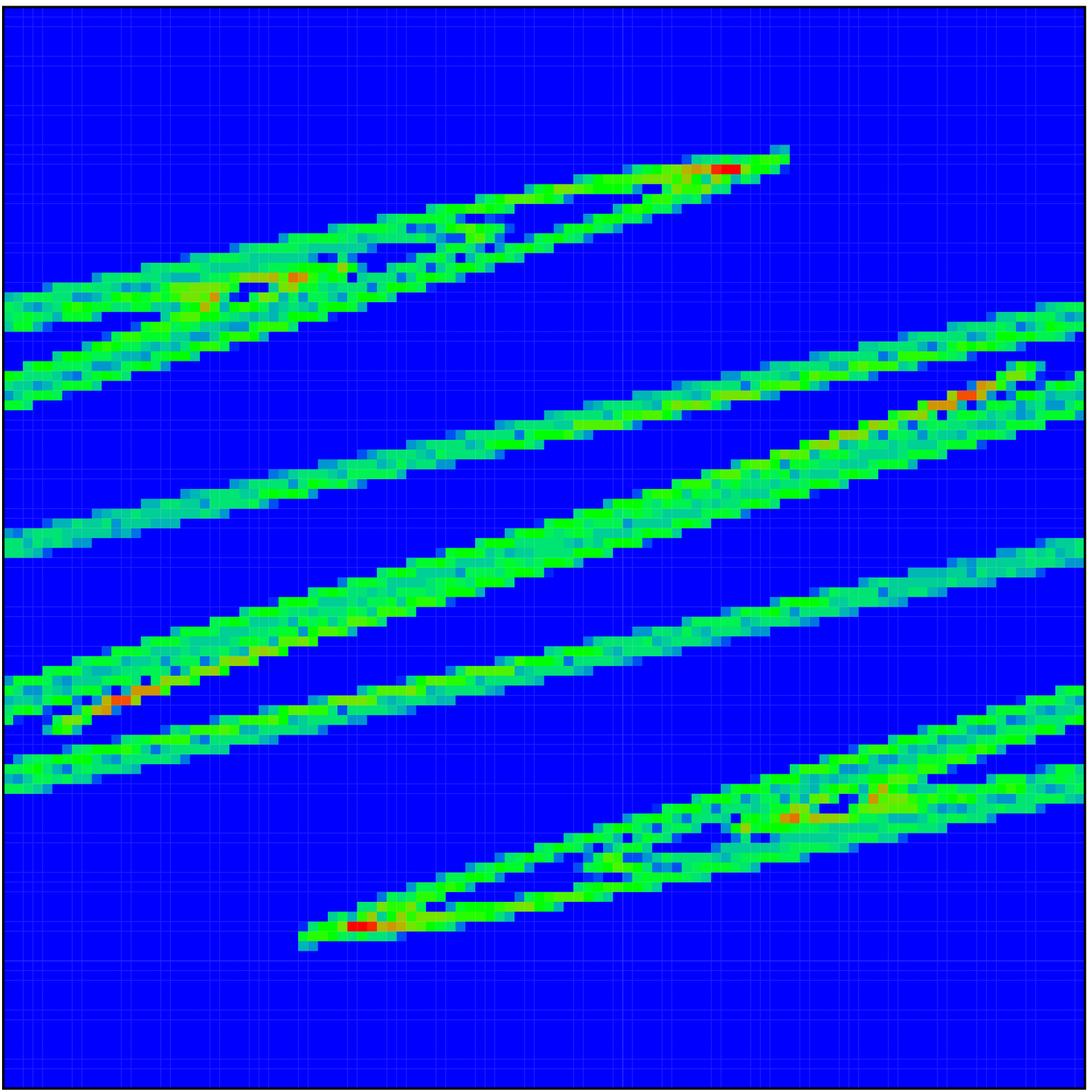} 
\caption{(Color online) Phase space representation of eigestates
$\psi_i$ of the Ulam matrix 
approximant  ${\bf S}$ of the Perron-Frobenius operator
for models 1 and 2 at $N=110 \times 110$
(color is proportional to $|\psi_i|$ with red/gray for maximum
and blue/black for zero). Left column shows eigenstates
for the model 1 at $K=7, a=2$ for maximum $\lambda_1 = 0.756$ (top panel)
and $\lambda_3 = -0.01+\imath0.513$ (bottom panel),
the space region is ($-aK/2 \leq y \leq aK/2, 0 \leq x \leq 2\pi$)
and the fractal dimension of the strange repeller is
$d=1.769$.
Right column shows eigenstates for the model 2 at
$K=7, \eta =0.3$ for maximum $\lambda=1$ (top panel)
and $\lambda_3 = -0.258+\imath0.445$  (bottom panel),
the space region is ($-4\pi \leq y \leq 4\pi, 0 \leq x \leq 2\pi$)
and the fractal dimension of the strange attractor is
$d=1.532$.}
\label{fig1}
\end{center}
\end{figure}

The fractal Weyl law gives the following scaling for the number of
Gamow states $N_\gamma$ with the decay rate in a finite band width 
$0 \leq \gamma \leq \gamma_b$:
\begin{equation}
N_\gamma \propto N^{\nu} \; , \;\; N=V/\hbar \;, \;\; \nu=d-1 \; ,
\label{eq1} 
\end{equation}
where $N$ is a matrix size given by a number of quantum states
in a volume $V$ and the exponent $\nu$
is determined by a fractal dimension $d$ of 
classical set formed by classical trajectories
nonescaping in future times (see Fig.1).

In view of the result (\ref{eq1}) it is natural to assume that
the fractal Weyl law should also work for other 
type of nonunitary matrix operators.
An important type of such matrices is generated by the Ulam method \cite{ulam}
applied to the Perron-Frobenius operators of dynamical systems \cite{mbrin}.
The method is based on discretization of the phase space and construction of
a Markov chain based on probability transitions between such discrete cells
given by the dynamics. It is proven that for hyperbolic maps 
in one and higher dimensions
the Ulam method converges to the spectrum of continuous system \cite{li}. 
While the spectrum of such Ulam matrix approximant 
of continuous operator has been studied numerically
for various dynamical maps (see e.g \cite{tel} and Refs. therein)
the validity of the fractal Weyl law has not been investigated.
Mathematical results for the Selberg zeta function \cite{guillope}
indicate that the law (\ref{eq1}) should remain valid but, as we  show here,
for certain dynamical systems the exponent $\nu$ starts to depend on 
fractal dimension $d$ in a different way.

It is known that in certain cases the Ulam method
gives significant modifications of the spectrum compared to
the case of the continuous Perron-Frobenius operators \cite{li}.
In fact discretization by phase-space cells
effectively introduces small noise added to dynamical equations of motion. 
For Hamiltonian systems
with divided phase space this noise destroys the invariant curves
and drastically changes the eigenstate of the Perron-Frobenius operator
(see e.g. discussion in \cite{dlszhirov}). However, for 
homogeneously chaotic systems the effect of this noise is rather
weak compared to dynamical chaos and thus , in the limit of small cell size,
the physical properties of the dynamics are expected to 
have no significant modifications in agreement with the results
presented in \cite{li,dlszhirov}.
Our numerical results obtained for 
dynamical maps with homogeneous chaotic dynamics 
confirm the convergence of the Ulam method to the
continuous limit of the Perron-Frobenius operator.

The paper is organized in the following way: Section II
gives the model description; Section III presents the numerical results
and the discussion is given in Section IV. 

\section{Model description}
To study the validity of the fractal Weyl law we use the Chirikov standard map
\cite{chirikov1979}. We consider two models: the map with absorption 
that corresponds to the classical limit of
the quantum model studied in \cite{borgonovi,dls2008} (model 1)
and the map with dissipation (model 2) also known as the Zaslavsky map 
\cite{zaslavsky}. In the first model the dynamics is described by the map
\begin{equation}
\left\{
\begin{array}{lll}
\bar{y} &=& y+K\sin(x+y/2) \\
\bar{x} &=& x+(y+\bar{y})/2 \ \ \ (\mathrm{mod} 2\pi)
\end{array}
\right.
\label{eq2}
\end{equation}
where bar notes the new values of dynamical variables and
$K$ is the chaos parameter. The map is written in its symmetric form
and all orbits going out of the interval $-aK/2 \leq y \leq aK/2$ are 
absorbed after one iteration. We consider a strong chaos regime at 
fixed $K=7$ and vary the classical escape time by changing
$a$ in the interval $0.8 \leq a \leq 6$. 

The second model is described by the map with dissipation parameter $\eta <1$:
\begin{equation}
\left\{
\begin{array}{lll}
 \bar{y} &=& \eta y + K \sin{x} \\
 \bar{x} &=& x +\bar{y}  \ \ \ (\mathrm{mod} 2\pi)
\end{array}
\right.
\label{eq3}
\end{equation}
with periodic boundary conditions in $y\in[-4\pi,4\pi)$. Due to dissipation
and chaos the dynamics converges to a strange attractor (see e.g. \cite{ott}).

To construct the Ulam matrix 
approximant for a continuous Perron-Frobenius operator 
in the two-dimensional phase space we divide the space of dynamical variables
$(x, y)$ on $N = N_x\times N_y$ cells with $N_x = N_y$. 
Then $N_c$ trajectories are propagated on one map iteration
from a cell $j$, and 
the elements $S_{ij}$ are taken to be equal to a relative number $N_i$ 
of trajectories arrived at a cell $i$ ($S_{ij} = N_i /N_c$ and 
$\sum_iS_{ij}=1$). 
Thus the matrix ${\bf S}$ gives a coarse-grained approximation of 
the Perron-Frobenius operator for the dynamical map. The map gives
about $K$ links for each cell.
We use $N_c$ values from $10^4$ to $10^6$  where the results are
independent of $N_c$. The fractal dimension $d$ of the strange repeller
and attractor depends on system parameters and is computed
as a box counting dimension using standard methods \cite{ott}.

We also used another method to construct the Ulam matrix based on a one trajectory
for the dynamics with a strange attractor in the model 2.
In the one trajectory Ulam method we iterate one trajectory up to time $t_1=100$;
after that we continue  iterations of the trajectory up to time $t=10^9$
and determine the matrix elements $S_{ij}$ as the ration between the number of transitions
from cell $j$ to cell $i$ divided by the total number of transitions $N_c$ from cell $j$
to all other cells (in this way $\sum_i S_{ij}=1$). This approach has certain advantages since it gives 
the Ulam matrix restricted to a dynamics only on the attractor. 
For a given cell size this method gives a significantly smaller matrix size $N_a \ll N$
since the number of cells $N_a$ located on the attractor is much smaller than the total number of cells $N$. 
When speaking about the results 
based on the one trajectory Ulam method we always directly specify this.

\section{Numerical results}

\begin{figure}[ht]
\centerline{\epsfxsize=7.2cm\epsffile{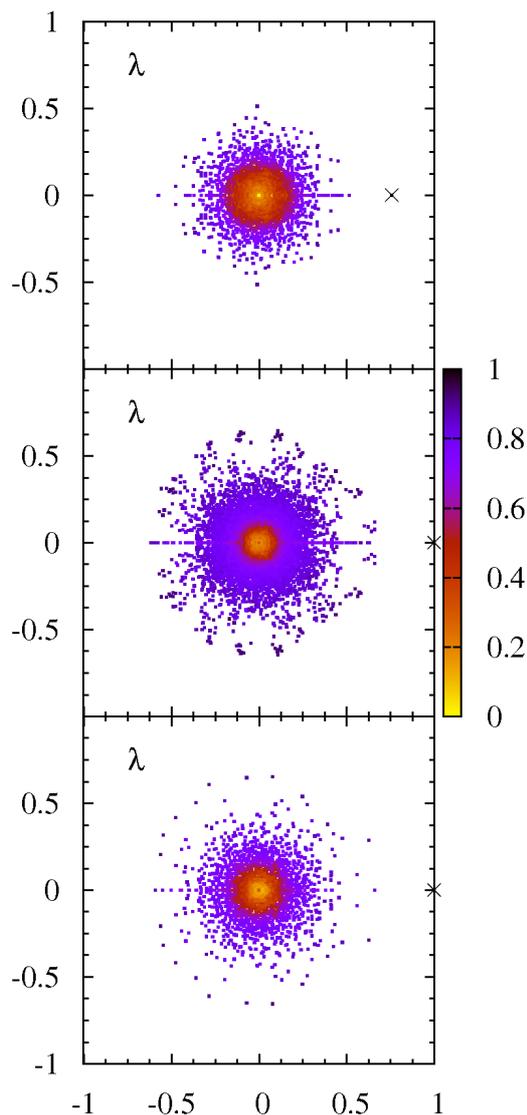}}
\caption{(Color online) Distribution of eigenvalues $\lambda$ in 
the complex plane for the Ulam matrix approximant 
${\bf S}$ for the parameters of Fig.1
for the models 1 (top panel) and 2 (center panel). 
Bottom panel shows the spectrum for the model 2,
with the same parameters as for the central panel,
obtained via the one trajectory Ulam approximant (see text). 
Color/grayness of small squares is determined 
by the value of overlap measure $\mu$ defined in the text
and shown  in the palette.}
\label{fig2}
\end{figure}

The eigenvalues $\lambda_i$ and right eigenvectors $\psi_i$
of the matrix ${\bf S}$ (${\bf S} \psi_i = \lambda_i \psi_i$)
are obtained by a direct diagonalization. Examples of the eigenstates
with maximal absolute values of $\lambda_i$ are shown in Fig.1.
The fractal structure of eigenstates is evident.
For the model 1 the measure is decreasing due to absorption
and $\lambda_1<1$, the state with $\lambda_1$
represents a set of strange repeller formed by orbits
nonescaping in future. For the model 2 all measure drops on 
the strange attractor and in agreement with the Perron-Frobenius theorem 
we have $\lambda=1$ \cite{mbrin}. Other eigenstates with smaller values
of $|\lambda|$ are located on the same fractal set as
the states with  maximal $\lambda_1$ but have another density distribution
on it.

The spectrum of matrix ${\bf S}$ in the complex plane 
is shown in Fig.2. It has a maximal real
value $\lambda_1$ isolated by a gap from a cloud of
eigenvalues more or less homogeneously distributed
in a circle of radius $r_\lambda$. For the model 2 the dense part
of the spectrum has $r_\lambda \approx \eta$ 
(at least at small values of $\eta$) that physically corresponds to
the fact that $\eta$ gives the relaxation rate to the limiting set
of the strange attractor. The gap between $\lambda_1$ 
and other eigenvalues in the model 1 is probably related to
a dynamics on the strange repeller. According to \cite{dls2008}
the decay rate of total probability in (\ref{eq2})
is exponential in time with the rate $\gamma_c=0.270$
(for parameters of Figs.1,2). This agrees well with
the numerical value $\lambda_1=0.756 \approx \exp(-\gamma_c)$.
The data of Fig.1 indicate that the states with $i>1$ 
have a strong overlap
with the steady state of $\lambda_1$ ($i=1$). 
In a quantitative way this overlap can be characterized 
by an overlap measure defined as
$\mu_i=\sum_l \psi_1(l) |\psi_i(l)|$ where the sum runs over
all $N$ cells. For $\mu$ close to unity
an eigenstate $\psi_i$ has a strong overlap
with the steady state $\psi_1$ and such states can be viewed
as higher mode excitations on this domain. For $\mu \ll 1$
we have other type of states being rather different from $\psi_1$.
The data of Fig.2 show that states with 
small values of $|\lambda|$ have small $\mu$. 

\begin{figure}[ht!]
\begin{center}  
\includegraphics[width=.48\textwidth]{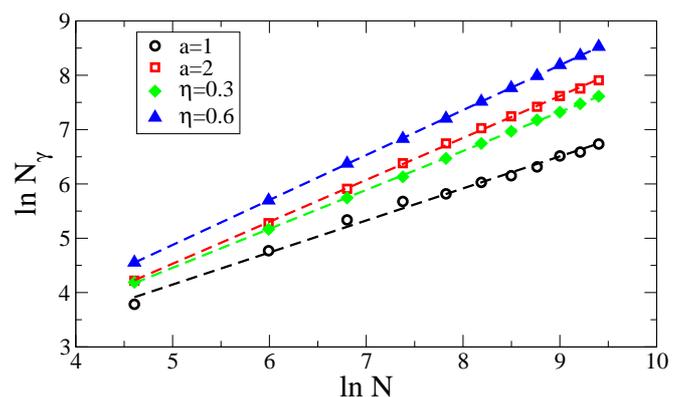} 
\caption{(Color online) 
Dependence of the integrated number of states $N_{\gamma}$ with decay rates 
$\gamma\leq\gamma_b=16$ on the size $N$ of the Ulam matrix ${\bf S}$
for the models 1 and 2 at $K=7$. The fits of numerical data, 
shown by dashed straight lines, give
$\nu=0.590, d=1.643$ (at $a=1$);
$\nu=0.772, d=1.769$ (at $a=2$);
$\nu=0.716, d=1.532$ (at $\eta=0.3$);
$\nu=0.827, d=1.723$ (at $\eta=0.6$).}
\label{fig3}
\end{center}
\end{figure}

In fact, as it is typical of the fractal Weyl law, almost all
eigenvalues drop to very small $|\lambda| \rightarrow 0$.
The number of states within a finite band with $0 \leq \gamma \leq \gamma_b$,
where $|\lambda|=\exp(-\gamma/2)$, grows algebraically with $N$
with the exponent $\nu <1$ remaining small compared to $N$. 
Typical examples of such a dependence are shown in Fig.3 for both models.

The spectrum for the one trajectory Ulam approximant is shown in the bottom panel of Fig.2
(here $N_a=2308$ while $N=12100$). The spectrum with one trajectory has a structure
similar to the spectrum of the usual Ulam method. This shows that 
the spectrum is mainly determined by the diffusive type excitations on the attractor.

\begin{figure}[ht!]
\begin{center}  
\includegraphics[width=.48\textwidth]{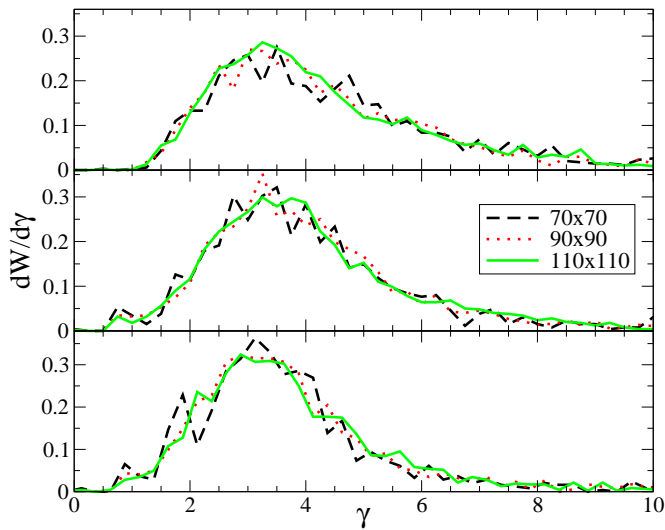} 
\caption{(Color online) Dependence of density of states $dW/d\gamma$ 
on the decay rate $\gamma$ for the Ulam matrix ${\bf S}$ 
for the  model 1 (top panel), and model 2
(center panel) at different 
sizes $N=N_x\times N_y$ given in the inset. 
Bottom panel shows the spectral density for the one trajectory Ulam method
for parameters of the central panel.
Data are shown for parameters of Fig.1, the density is normalized
by the condition $\int_0^{16} dW/d \gamma d \gamma=1$.}
\label{fig4}
\end{center}
\end{figure}

Our matrix sizes $N$ are sufficiently large and allow to reach
asymptotic behavior 
in the limit of large $N$. This is confirmed by the fact that the
density of states $dW/d\gamma$ in $\gamma$ becomes 
independent of $N$ as it is show in Fig.~4. This directly demonstrates that the Ulam method
is stable for our models and that it converges to the continuous limit
of the Perron-Frobenius operator. The density of states for the Ulam matrix obtained with
one trajectory has the density of states very close to the one obtained by the
usual Ulam method. This shows that the spectrum with finite values of $\gamma$
is determined by the dynamics on the attractor.

The density is 
mainly determined by the
cloud of states in the radius $r_\lambda$, the contribution of the
isolated eigenvalue $\lambda_1$ is only weakly visible at minimal
$\gamma$. The density has a broad maximum around
$\gamma \approx 3$, for the model 2 this value is compatible with
the value $- 2 \ln \eta$ which determines the global relaxation rate to 
the strange attractor. It is interesting to note that for the model 1
the spectral density of the Perron-Frobenius operator (Fig.4, top panel) is 
rather different from the spectral density 
in the corresponding quantum problem (Fig.4 in \cite{dls2008}).
Indeed, the densities $dW/d\gamma$ for the classical and quantum systems
are very different: the classical model 1 has one isolated
eigenvalue $\lambda_1$ and a broad maximum around $\gamma \approx 3$.
The quantum model of \cite{dls2008} has a peaked distribution
around $\gamma_c=-2\ln \lambda_1$ corresponding to the classical state
at $\lambda_1$ and a monotonically 
decreasing density at larger values of $\gamma$. 
At the same time the eigenstates with minimal $\gamma$
are located on the strange set of trajectories nonescaping in future times, 
both in the classical and quantum cases
(see Fig.1 here and in \cite{dls2008}). Thus the semiclassical
correspondence between classical and quantum cases of model 1
still requires a better understanding.

\begin{figure}[ht!]
\begin{center}  
\includegraphics[width=.48\textwidth]{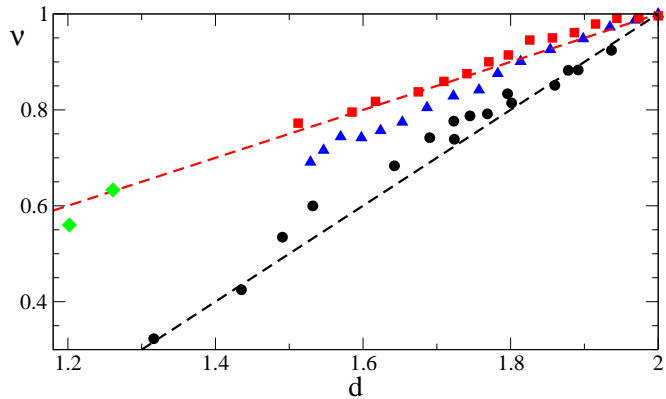}
\caption{(Color online) Fractal Weyl law for three different models: 
models 1 (black points at $K=7$) and 2 (red/gray squares at $K=12$,
blue/black triangles at $K=7$) and H\'enon map 
(green/gray diamonds at $a=1.2; 1.4$ for $b=0.3$). 
The fractal Weyl law exponent $\nu$ is shown as a function
of fractal dimension $d$ of the strange forward trapped set in model 1
and strange attractor in model 2 and Henon map.
The straight dashed lines show the laws (\ref{eq4}) (upper line)
and (\ref{eq1}) (bottom line).
We used $a\in[0.8,6]$ for model 1 and 
$\eta\in[0.3,1]$ for model 2.}
\label{fig5}
\end{center}
\end{figure}

We determine the exponent $\nu$ as it is shown in Fig.~\ref{fig3}
for both models at different values of parameters.
At the same time we compute the fractal dimension $d$ of the
strange set of trajectories nonescaping in future  using box counting
dimension with a box size $\epsilon$. 
In this way the size of the Ulam matrix is $N=1/\epsilon^2$
while the number of cells on the fractal set
scales as $N_f \propto 1/\epsilon^d = N^{d/2}$.
In this way we determine the dependence of $\nu$ on $d$.
The data are shown in Fig.~5. For the model 1 we
find that the usual fractal Weyl law with
$\nu=d-1$ holds in a large interval of variation of $d$.
Relatively small deviations can be attributed to a finite
accuracy in computation of $\nu$ at finite matrix sizes.
In contrast to that for the model 2 we find absolutely
another relation which can be approximately described as 
\begin{equation}
 N_\gamma \propto N^{\nu} \; , \;\; \nu=d/2 \; .
\label{eq4}
\end{equation}

This relation works rather well for $K=12$ while for
$K=7$ the deviations are a bit larger. We attribute this
to the fact that at $K=7$ there is a small island of stability
at $\eta=1$ \cite{chirikovk7} which does not influence the
dynamics in the case of absorption (\ref{eq2}) but can produce 
certain influence for the dissipative case (\ref{eq3}).
To check that the law (\ref{eq4}) works for other systems 
with strange attractors we computed $\nu$ and $d$ for the H\'enon
map (${\bar x}=y+1 -a x^2, {\bar y}=b x$, see e.g. \cite{ott}) at 
standard parameter values of $a, b$. The results
confirm the validity of the fractal Weyl law
also for the H\'enon map (see Fig.~5).

The physical origin of the law (\ref{eq4})
can be understood in a simple way: the number of states $N_\gamma$
with finite values of $\gamma$ is proportional to the number of
cells $N_f \propto N^{d/2}$ on the fractal set of strange attractor.
Indeed, the results for the overlap measure $\mu$ (see Fig.~2)
show that these states have strong overlap with the
steady state while the states with $\lambda \rightarrow 0$
have very small overlap. Thus almost all $N$
states have eigenvalues  $\lambda \rightarrow 0$
and only a small fraction of states on the strange attractor
$N_\gamma \propto N_f \propto N^{d/2} \ll N$
has finite values of $\lambda$. We also checked that the participation
ratio $\xi$ of the eigenstate of model 2 at $\lambda=1$,
defined as $\xi=(\sum_l |\psi_1(l)|^2)^2/\sum_l |\psi_1(l)|^4$,
grows as $\xi \sim N_f \propto N^{d/2}$.

The fractal Weyl laws (\ref{eq1}) and (\ref{eq4}) have
two different exponents $\nu$ but they correspond to two
different situations: for (\ref{eq1}) the law describes
the systems with absorption when all measure
escapes from the system and only a small fractal set
remains inside; for  (\ref{eq4}) all measure drops on a
fractal set inside the system. Due to that reasons the exponents
are different.

\begin{figure}[ht!]
\begin{center}  
\includegraphics[width=.48\textwidth]{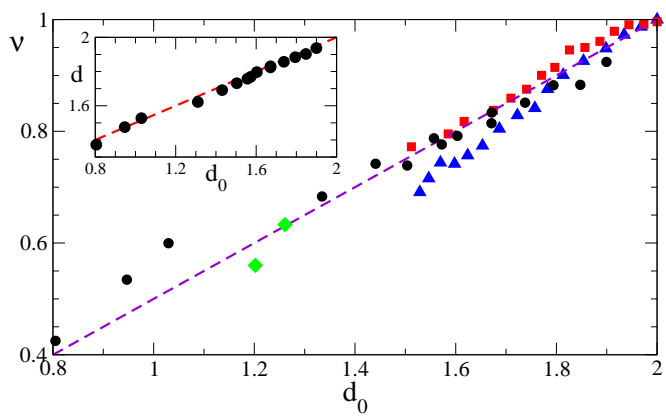}
\caption{(Color online) 
Fractal Weyl law for three different models as a function of the dimension 
of the invariant set $d_0$; the models and their parameters are the same 
as in Fig. \ref{fig5}. The fractal Weyl exponent $\nu$ is shown as a function
of fractal dimension $d_0$ of the strange repeller in model 1
and strange attractor in model 2 and Henon map.
The straight dashed line show the theoretical dependence $\nu=d_0/2$ 
of Eq.~(\ref{eq5}).
The inset shows the relation between the fractal dimension $d$ of 
 trajectories nonescaping in future
and the fractal repeller dimension $d_0$ for  the case of model 1;
the dashed straight line shows the theoretical dependence 
$d=d_0/2+1$.
}
\label{fig6}
\end{center}
\end{figure}

The different dependencies  of $\nu$ on $d$ in Eqs.~(\ref{eq1},\ref{eq4}) 
can be reduced to 
one dependence if to express $\nu$ via the fractal dimension $d_0$ of the
invariant sets. Indeed, for the model (2) all trajectories 
drop on the strange attractor which can be considered 
as an invariant set with the fractal dimension $d_0=d$.
For the model 1 we have the set of trajectories nonescaping in future
with dimension $d$, there is also the fractal set of trajectories nonescaping in the past
which has also the dimension $d$  due to symmetry between the future and the past
present in the model 1 
(symmetry to reflection $x,y \rightarrow -x, -y$ in  (\ref{eq2})).
Then the invariant set of a strange repeller
corresponds to the intersection of these two sets of trajectories nonescaping neither in 
the future neither in the past with the fractal dimension $d_0$.
As it is known, see e.g. \cite{ott}, we have
$2=d+d-d_0$ so that $d=d_0/2+1$. This relation is confirmed by the data
presented in the inset of Fig.~6. On the basis of these relations we can express the fractal Weyl
exponent via the fractal dimension $d_0$ of the invariant set 
\begin{equation}
 N_\gamma \propto N^{\nu} \; , \;\; \nu=d_0/2 \; .
\label{eq5}
\end{equation}
This global dependence is confirmed by the data shown in Fig.~6.

The numerical data for the one trajectory Ulam method gives
always $N_\gamma \propto N_a$. This satisfies the relation
(\ref{eq5}) since by definition $N_a \propto N^{d_0/2}$.

\section{ Discussion}
In summary, our results show that the Ulam method gives 
very efficient possibility to study the spectral 
properties of the Perron-Frobenius operators
for systems with dynamical chaos. Their spectrum 
is characterized by  the fractal Weyl law
with the Wyel exponent determined by the fractal dimension
of dynamical system according to relations
(\ref{eq1}), valid for systems with absorption or chaotic scattering,
 or (\ref{eq4}), valid for dissipative systems with strange attractors.

It is interesting to note that for dynamical systems
the Ulam method naturally generates  directed Ulam networks
\cite{dlszhirov} which have certain similarities with 
the properties of the Google matrix of the World Wide Web (WWW).
However, for  the model 2 and the H\'enon map considered above, there is 
a finite gap between $\lambda=1$ and other eigenvalues
while for the WWW there is no such gap \cite{brin,googlebook}.
In this sense the above models are more close to randomized 
directed networks considered in \cite{ggs}
which have a relatively large gap. 
We note that the PageRank vector $\psi_1$ with $\lambda=1$,
used by Google for ranking of web pages,
corresponds in our case to a strange attractor.
In this case the probability $p_l \sim \psi_1(l)$
is distributed over all cells $N_f \propto N^{d/2}$ occupied
by the strange attractor. The number  of such cells
grows infinitely with $N$ that corresponds to a delocalized
phase of the PageRank similar to the cases discussed in \cite{dlszhirov}.
In contrast to that the WWW is characterized by a localized PageRank
with an effective finite number of
populated sites independent of $N$. In spite of that it is not excluded
that the future evolution of the WWW can enter in 
a delocalized regime of the PageRank. Therefore,
we think that the fractal Weyl law 
discussed here can be useful not only for 
the Perron-Frobenius operators of dynamical systems but also
for various types of realistic directed networks.

\end{document}